\documentclass[ai,preprints,accept,moreauthors,pdftex]{Definitions/mdpi} 

\firstpage{1} 
\pubvolume{1}
\issuenum{1}
\articlenumber{0}
\pubyear{2021}
\copyrightyear{2020}
\datereceived{} 
\dateaccepted{} 
\datepublished{} 
\hreflink{https://doi.org/} 

\Title{Assessing glaucoma in retinal fundus photographs using Deep Feature Consistent Variational Autoencoders}

\TitleCitation{Assessing glaucoma in retinal fundus photographs using Deep Feature Consistent Variational Autoencoders}

\Author{Sayan Mandal $^{1,2}$\orcidA{}, Alessandro A. Jammal $^{2}$\orcidB{} and Felipe A. Medeiros $^{1,2,}$*}

\AuthorNames{Sayan Mandal, Alessandro Jammal and Felipe Medeiros}

\AuthorCitation{Mandal, S.; Jammal, A.A.; Medeiros, F.A.}

\address{%
$^{1}$ \quad Department of Electrical and Computer Engineering, Pratt School of Engineering, Duke University, Durham, NC; \\
$^{2}$ \quad Vision, Imaging, and Performance Laboratory (VIP), Duke Eye Center and Department of Ophthalmology, Duke University, Durham, NC.}

\corres{Corresponding author: Felipe A. Medeiros, MD, PhD, Duke Eye Center, Department of Ophthalmology, Duke University, 2351 Erwin Rd, Durham, North Carolina, 27705. E-mail: felipe.medeiros@duke.edu}

\abstract{One of the leading causes of blindness is glaucoma, which is challenging to detect since it remains asymptomatic until the symptoms are severe. Thus, diagnosis is usually possible until the markers are easy to identify, i.e., the damage has already occurred. Early identification of glaucoma is generally made based on functional, structural, and clinical assessments. However, due to the nature of the disease, researchers still debate which markers qualify as a consistent glaucoma metric. Deep learning methods have partially solved this dilemma by bypassing the marker identification stage and analyzing high-level information directly to classify the data. Although favorable, these methods make expert analysis difficult as they provide no insight into the model discrimination process. In this paper, we overcome this using deep generative networks, a deep learning model that learns complicated, high-dimensional probability distributions. We train a Deep Feature consistent Variational Autoencoder (DFC-VAE) to reconstruct optic disc images. We show that a small-sized latent space obtained from the DFC-VAE can learn the high-dimensional glaucoma data distribution and provide discriminatory evidence between normal and glaucoma eyes. Latent representations of size as low as 128 from our model got a 0.885 area under the receiver operating characteristic curve when trained with Support Vector Classifier. 
}

\keyword{Deep Generative Networks; Deep Feature Consistent Variational Autoencoders; Support Vector Classifier, Optic Disc Images, Glaucoma} 

\begin{document}

\section{Introduction}

Glaucoma is an optic neuropathy that is associated with progressive vision loss. It is one of the leading causes of blindness, and it is estimated that 111.8 million people worldwide will be affected by the disease by 2040 \cite{mariotti2012global, global2014tham}. The current standard for diagnosing glaucoma is based on a computerized assessment of the patient’s visual field and the observation of anatomical changes in the optic disc and retinal nerve fiber layer (RNFL) \cite{jonas1999ophthalmoscopic, medeiros2009prediction}. Although prominent, the disease remains asymptomatic until late stages, when central vision and visual acuity may be irreversibly lost. In contrast, progressive optic disc damage may appear several years before detectable substantial changes are seen on standard perimetric tests and pose an opportunity for early disease detection \cite{li2020impact,kass2002ocular,miglior2005}. 

In medical imaging, retinal fundus photography and, more recently, optical coherence tomography (OCT) have been widely used to document structural damage to the optic disc from glaucoma. In particular, OCT has gained popularity due to its ability to detect even the most minute changes in the RNFL thickness with high reproducibility \cite{medeiros2021}. However, the optimal use of OCTs for glaucoma is highly dependent of well-trained technicians for image acquisition and its use is limited due to high operation costs \cite{siam2008}. Retinal fundus photographs, on the other hand, have attractive features for glaucoma detection due to various factors, such as low cost, compact size of devices, and low requirement for operating skills \cite{retinal2010abramoff}.

In recent years, various deep learning (DL) algorithms have been introduced using fundus photographs, which alleviates reproducibility and performance issues reported previously with manual review \cite{expert1992varma, intraobserver1988tielsch, medeiros2021}. Innovative ground truth generation methods such as OCT-acquired RNFL thickness measurements for DL have improved these shortcomings. Dubbed Machine to Machine (M2M), this technique has not only enhanced glaucoma diagnostic accuracy with fundus photos \cite{LEE202186} but also provided a quantitative estimates that can be used to track progression \cite{detection2021shabbir}. Despite these developments, standard DL methods on fundus photos have limitations, such as variability in the quality of the images \cite{LEE202186}, underlying data distribution \cite{LEE202186,detection2021shabbir}, and most importantly, interference due to non-glaucoma prominent features (different eye diseases) \cite{computer2018hagiwara}, which may cause hindrance in classification.

One way to approach these issues is by training robust DL models on representative datasets (usually requiring abundant high-quality data). However, since common DL approaches are based on supervised learning, such models are dependent on limited labels and incomplete data distributions. Unsupervised learning, especially deep generative models, alleviates this issue by learning unknown patterns in the data. Specifically, models like Variational Autoencoders (VAE) and Generative Adversarial Networks (GAN) tend to capture the approximate data distribution by reconstructing the input \cite{kingma2013auto, goodfellow2020generative}. Previous research has also shown that autoencoder based reconstruction scan capture novel patterns in different medical data analysis tasks \cite{tour2021raza}.

In the present work, we propose a VAE based analysis for glaucoma assessment in optic discs obtained from retinal fundus photographs. Specifically, we develop a Deep Feature Consistent VAE (DFC-VAE), which can preserve spatial correlation characteristics to reconstruct optic disc images \cite{dfcvae}. The DFC-VAE results are assessed based on similarity scores and the differences between the original image and the reconstruction \cite{dfcvae}. We then check the robustness of the model by varying the DFC-VAE and observing the model’s ability to capture glaucoma-related low-dimensional features at different latent space sizes. A standard Linear Support Vector Classifier (SVC) is used to compare the Area under the Receiver Operating Characteristic (ROC) Curve scores of glaucoma classification by low-dimensional latent features \cite{Platt99probabilisticoutputs}.  We show that, the low-dimensional latent representations obtained from the trained DFC-VAE approximates relevant feature distribution responsible for glaucoma assessment.

\section{Materials and Methods}

\subsection{Dataset}
\label{dataset}
This study included a retrospective cohort extracted from the Duke Glaucoma Registry, a database of electronic medical records developed by the Visual, Imaging, and Performance Laboratory, Duke Eye Center, Durham, NC. \cite{jammal2021}. The database consisted of adults at least 18 years of age who were evaluated at the Duke Eye Center or its satellite clinics between 1994 and 2019. The Duke University Institutional Review Board approved this study with a waiver of informed consent due to the retrospective nature of this work. The study was conducted in alignment with the tenets of the Declaration of Helsinki and regulations of the Health Insurance Portability and Accountability Act.

The dataset consisted of optic disc images taken with the Nidek 3DX (Nidek, Gamagori, Japan) and the Visupac FF-450 (Carl Zeiss Meditec, Inc, Dublin, CA), clinical data (including follow-up, diagnosis, medical history, etc.), standard automated perimetry (SAP) (Humphrey Field Analyzer II, Carl Zeiss Meditec, Inc, Dublin, CA) data and Spectralis spectral-domain optical coherence tomography (SDOCT) (Heidelberg Engineering GmbH, Dossenheim, Germany) scans for normal eyes and eyes with open-angle glaucoma. Subjects were excluded if they had a history of ocular or systemic diseases that could affect the optic nerve, other than glaucoma. Tests performed after any diagnosis of retinal detachment, retinal or malignant choroidal tumors, non-glaucomatous disorders of the optical nerve and visual pathways, exudative, atrophic and late-stage dry age-related macular degeneration, amblyopia uveitis and venous or arterial retinal occlusion according to International Classification of Diseases (ICD) codes were excluded. In addition, tests performed after treatment with panretinal photocoagulation, according to Current Procedural Terminology (CPT) codes, were also excluded, as described previously \cite{jammal2021}.

\begin{figure}[H]
\centering
\includegraphics[width=0.48\textwidth]{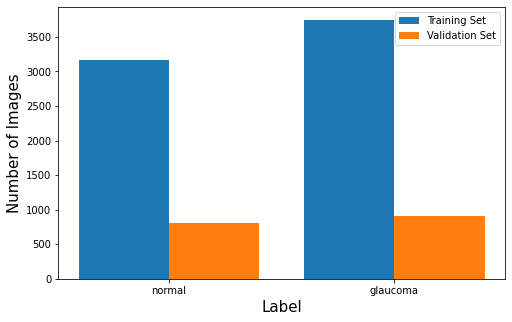}
\caption{Dataset distribution used for DFC-VAE based glaucoma assessment of retinal fundus photographs focused on the optic discs. \label{fig:split}}
\end{figure}

Included eyes were classified into glaucoma or normal based on a structure and function objective reference standard previously described \cite{mariottoni2021obj}, applied to the tests at baseline. In brief, the objective criteria for glaucomatous optic neuropathy accounts for the presence and correspondence of structural and functional defects, defined from parameters from SDOCT and SAP. To be considered glaucoma, an eye had to meet the criteria for global loss (i.e., global RNFL thickness from SDOCT outside normal limits and abnormal SAP, as defined by Glaucoma Hemifield Test (GHT) outside normal limits or pattern standard deviation (PSD) with $P < 5\%$); or localized loss (at least one sector in the RNFL thickness outside normal limits with a corresponding abnormality on the opposite SAP Hemifield, defined as a Hemifield Mean Deviation (MD) with $P < 5\%$). Eyes with normal SAP and SDOCT parameters were considered normal. SDOCT-SAP pairs that did not meet criteria for glaucoma or normal were considered suspects (i.e., only structural or only functional damage) and were not included in the analysis. The overall statistics of the dataset have 6902 image-label pairs (3158 - normal, 3744 -glaucoma) in the training set and 1725 image-label pairs (812 - normal, 913 - glaucoma) in the validation/test set (Figure~\ref{fig:split}).

\subsection{Variational-Autoencoder Model}
\label{vaesection}
\begin{figure}[H]
\centering
\includegraphics[width=0.35\textwidth]{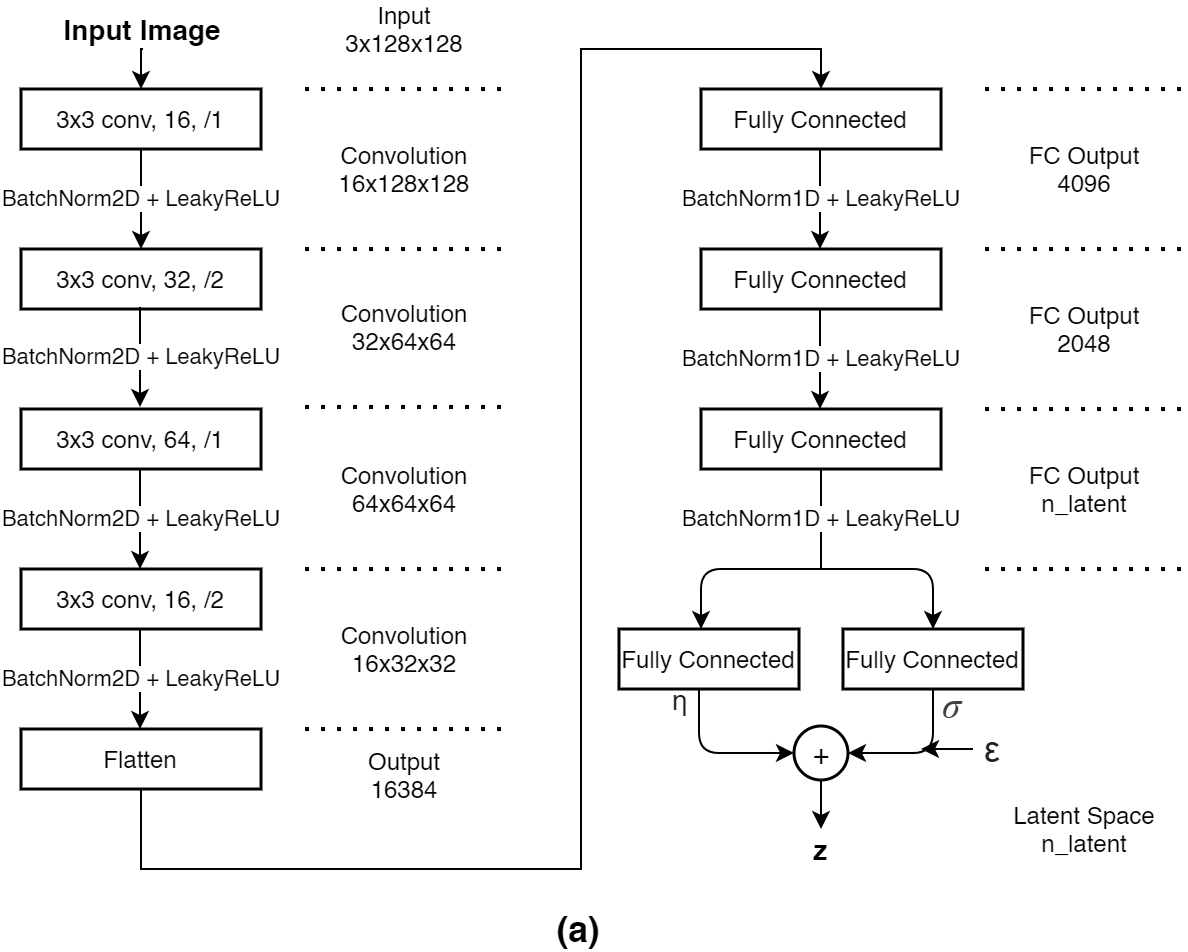}
\hspace{0.2cm}
\includegraphics[width=0.35\textwidth]{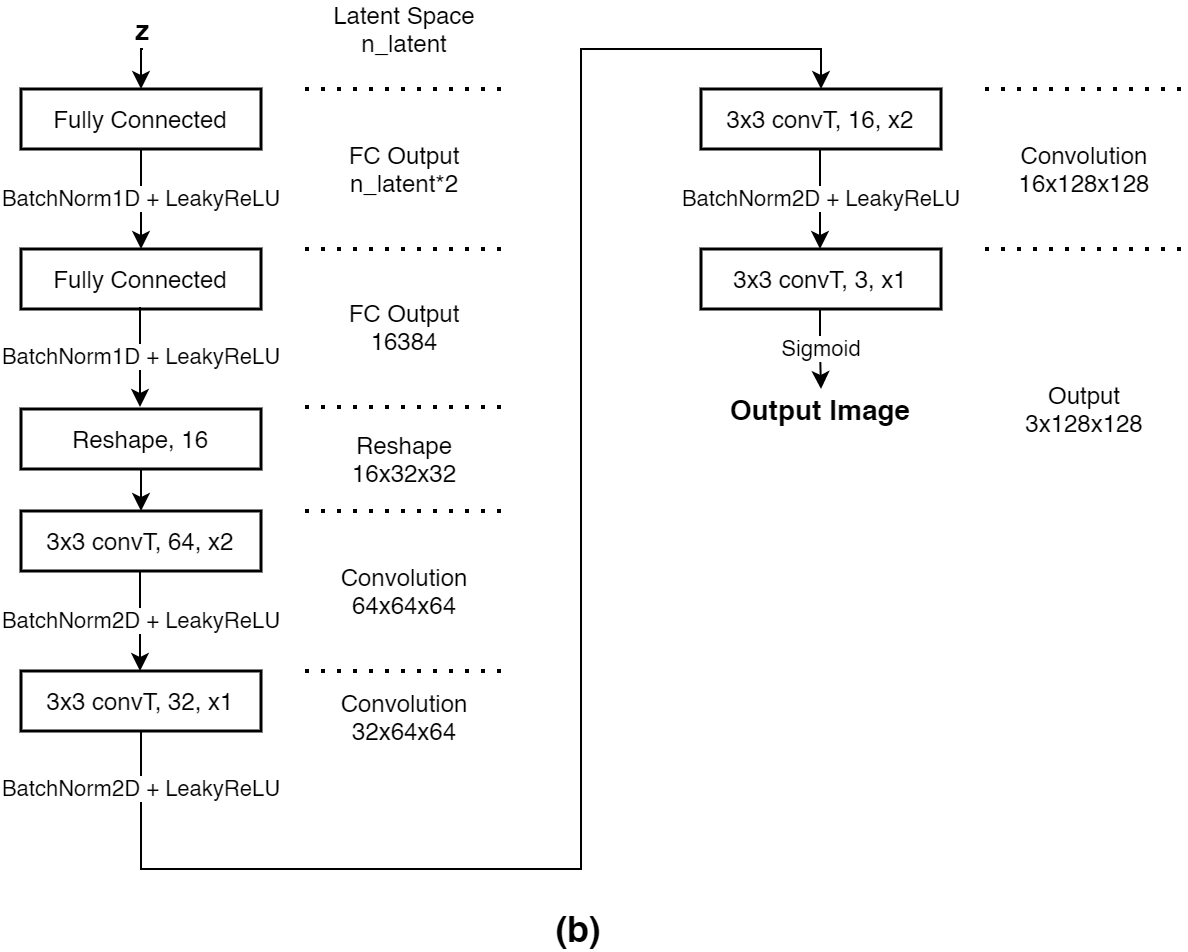}
\caption{Flow chart for DFC-VAE (\textbf{a}) encoder architecture and (\textbf{b}) decoder architecture. The DFC-VAE encoder takes optic disc image $x$ as input and compresses it in it's latent space $z$. The decoder uses the latent representations $z$ and generates new reconstructions $x'$  \label{fig:dfcvae}}
\end{figure}

The DFC-VAE model developed is similar to a VAE model, which can approximate data distribution for all observations and features in the input space. Since VAEs provide a probabilistic model to describe the data, they retain most of the information in the reduced representation (encoding) and generate new data by sampling it from the reduced representation distribution (decoding). In the VAE process, the input images (optic disc photos) $x$ are believed to come from some probability distribution. This data is stochastically encoded into a low-dimensional bottleneck, called latent representation $z$ using a gaussian probability distribution, $q_{\theta}(z|x)$ (encoder), where all possible data observations are encoded. The VAE then samples a point from the latent space and uses another Gaussian probability distribution, $p_{\phi}(x|z)$ (decoder) to generate a reconstruction, $x’$ whose properties match the original probability distribution of the input, $x$ \cite{kingma2013auto}. Since the input data type is RGB images, convolutional neural networks (CNN) are used in the earlier encoding phase to learn the local and global attributes of the images. Fully connected (FC) layers follow these layers for dimensionality reduction (Figure~\ref{fig:dfcvae}a). Similarly, FC layers are stacked with transposed convolution layers in the decoder to reconstruct all the spatial properties of the input data (Figure~\ref{fig:dfcvae}b).

\begin{figure}[H]
\centering
\includegraphics[width=0.35\textwidth]{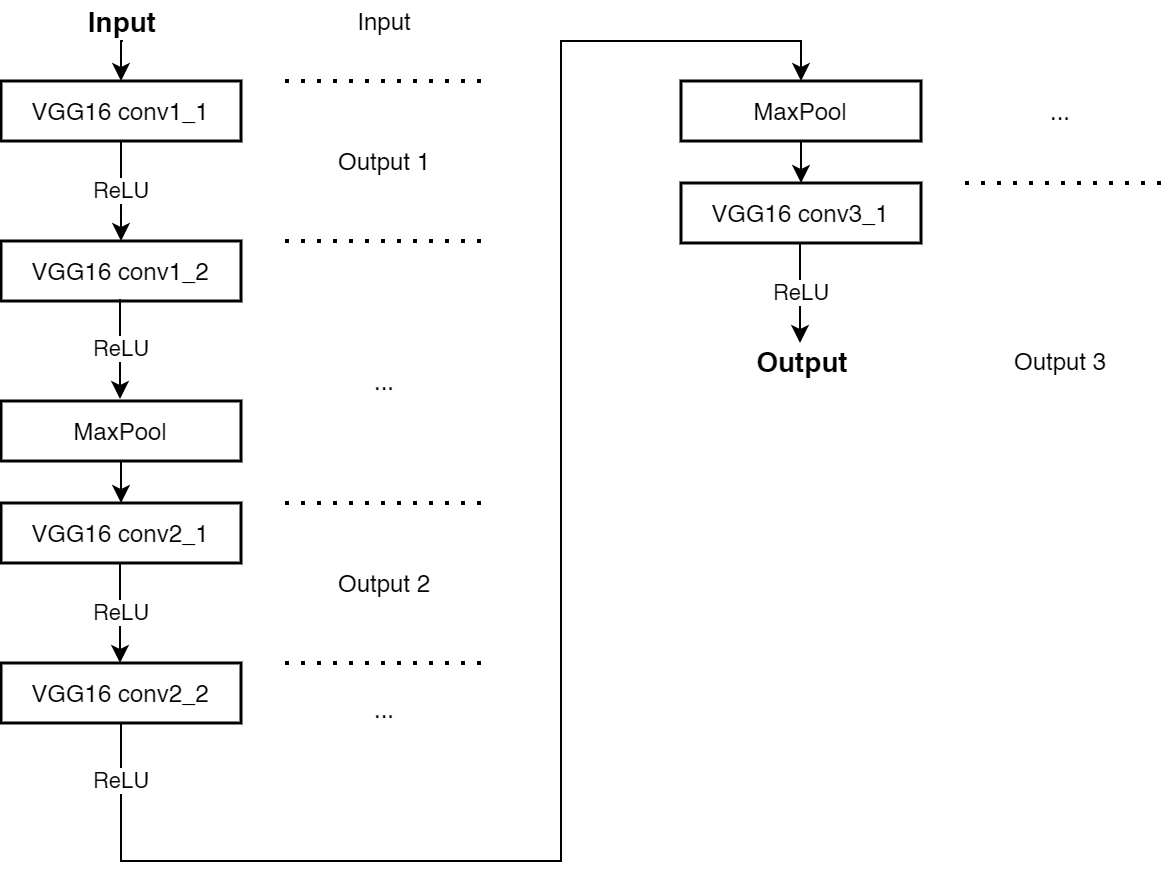}
\caption{Flow chart for hidden feature extraction architecture from a pretrained VGG16 model. For each input (original image or reconstruction), three different outputs are obtained at different hidden layers. \label{fig:vgg}}
\end{figure}

A regularized reconstruction error accompanies all VAE models, usually backpropagated through the VAE network. This error is made of a combination of pixel-by-pixel reconstruction loss and Kullback-Leibler (KL) divergence to ensure that most information is retained between the encoding and decoding phases. Since the higher quality and sharper edges of optic disc images aid in glaucoma analysis, preserving these information becomes of utmost importance. Therefore, replacing the spatially unaware pixel-by-pixel reconstruction loss with a more perceptual feature loss helps retain most information in optic disc images. This feature perceptual loss is derived from the difference between reconstruction and input hidden features in a pre-trained CNN model, usually taken at some hidden layer representations (1 or many) \cite{dfcvae}. This study uses the first three layers $l=1,2,3$ of VGG16 $\psi(x)$ (Figure~\ref{fig:vgg}) to extract the feature perceptual loss given by the equation\begin{equation}
\mathcal{L}_{fl} = \sum_{l = 1,2,3} \left( \frac{1}{2C^{l}W^{l}H^{l}} \sum_{c=1}^{C^l}\sum_{w=1}^{W^l}\sum_{h=1}^{H^l}\left(\psi(x)^l_{c,w,h} - \psi(x')^l_{c,w,h} \right)^2\right).
\end{equation}This process makes the VAE model perceptive to the perceptual difference between input and reconstruction while preserving most spatial correlations between them. Thus the whole process is called DFC-VAE \cite{dfcvae}.

\subsection{Training and Validation}
Optic disc images of the dataset were used as the input for the DFC-VAE training and evaluation. The 3-channeled RGB images were resized to $3 \times 128 \times 128$ pixels (49152 pixels), with a batch size of 64. As shown in section \ref{dataset}, the data split was kept at 0.2. In addition, a random horizontal flip with p = 0.5 was introduced at the training phase to increase the variability in the dataset. With the objective function discussed in section \ref{vaesection}, Adam optimizer with an initial learning rate of 1e-3 and scheduler with step size 140 and gamma 0.1 was used to train the DFC-VAE model for 300 epochs. Evaluation was done after every epoch, and the model with the lowest validation loss was saved. The whole process was assessed and repeated for different latent space sizes ranging from $2^1, 2^2, \dots , 2^{11}$.

The experimental setup used a stand-alone PC with a 32GiB Intel (R)  Core (TM) i7-8700K CPU, equipped with 12GiB Nvidia Geforce RTX 2080Ti GPU and Cuda 10.2 support. All codes for the DFC-VAE training process were written and executed in Pytorch, Python 3.8. Both image reconstructions and latent features served as the output of the trained DFC-VAE model, which were used for qualitative and quantitative analysis.

\subsection{Feature Selection and Clustering}
All evaluations were done on the validation dataset. The reconstructions obtained from the DFC-VAE were observed holistically by analyzing its structural similarity with the input. This analysis was done to verify if human graders can discern normal-glaucoma eyes at different latent space sizes. The smallest latent space size where this difference becomes apparent was selected, along with the largest latent space size, which preserves most information in the reconstructed image. Low-dimensional representations obtained from the DFC-VAE at these two latent space sizes were analyzed using clusters. Since DFC-VAE learns an approximate data distribution of the input, it may be hard to analyze how glaucoma-related features were learned. Top-k latent features of the low-dimensional representation which are most correlated with the objective criteria labels were ranked and selected to examine for glaucoma-related attributes. Finally, Uniform manifold approximation and projection (UMAP) learning, a dimension reduction algorithm, was employed for dimensionality reduction and clustering for qualitative analysis \cite{mcinnes2018umap}.  

\subsection{Glaucoma Classification Model}
A machine learning-based glaucoma classification model was used to show if the DFC-VAE model has observed glaucoma attributes in the optic disc images. The classification model was developed with latent representations of the DFC-VAE as input features and labels of the dataset as input labels.  A Support Vector Classifier (SVC), which separates the data by finding and projecting the input space on a hyperplane \cite{cortes1995support}, was used to classify the glaucoma eyes. This model was evaluated based on five performance metrics - accuracy, Area under the ROC Curve (AUC), F1 score, precision, and recall. Finally, results obtained from the classification model were used to validate the qualitative analysis done in the previous section.

DFC-VAE latent representations of the validation set (Figure~\ref{fig:split}) were used as the dataset for the classification model. Two types of analysis were done for classification: a 5-fold cross-validation model to obtain an optimal SVC classifier and a train-test split model to get the optimal latent space size that captures most glaucoma attributes. The performance of the 5-fold cross-validation model was done using all five metrics discussed above. The train-test split model used a split ratio of 0.3 with the optimal SVC as a classifier. This training was followed by an evaluation using the ROC curve and AUC scores for each latent space size.

\section{Results}

\begin{figure}[H]
\centering
\includegraphics[width=0.7\textwidth]{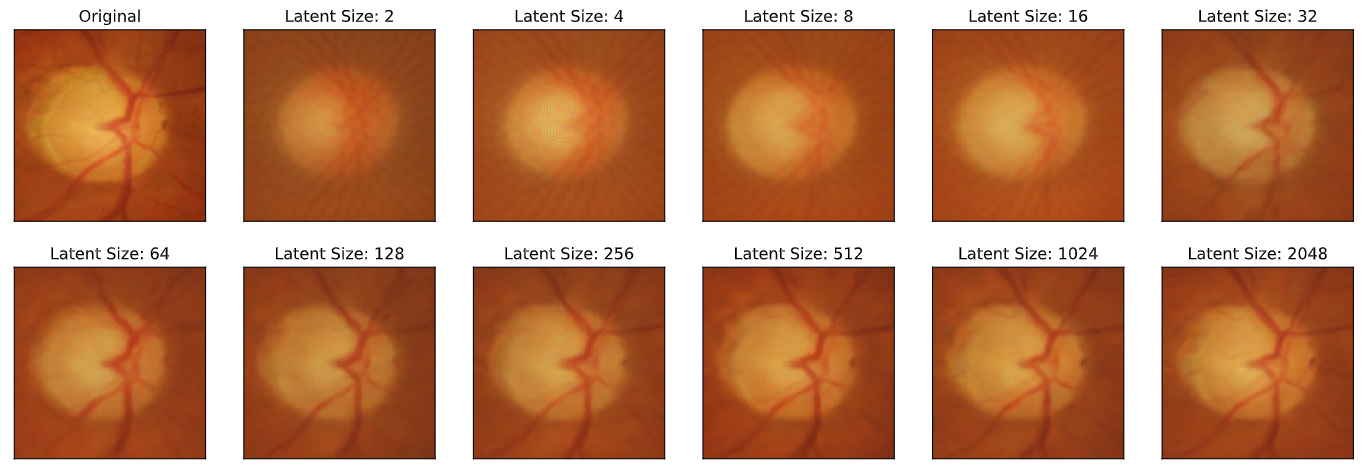}
\caption{Original followed by reconstruction images for optic disc with glaucoma obtained from trained DFC-VAE at different latent space sizes. The image shows reconstruction for optic disc at latent space sizes $nl \geq 128$ has discernible glaucoma traits. \label{fig:recglau}}
\end{figure}

\begin{figure}[H]
\centering
\includegraphics[width=0.7\textwidth]{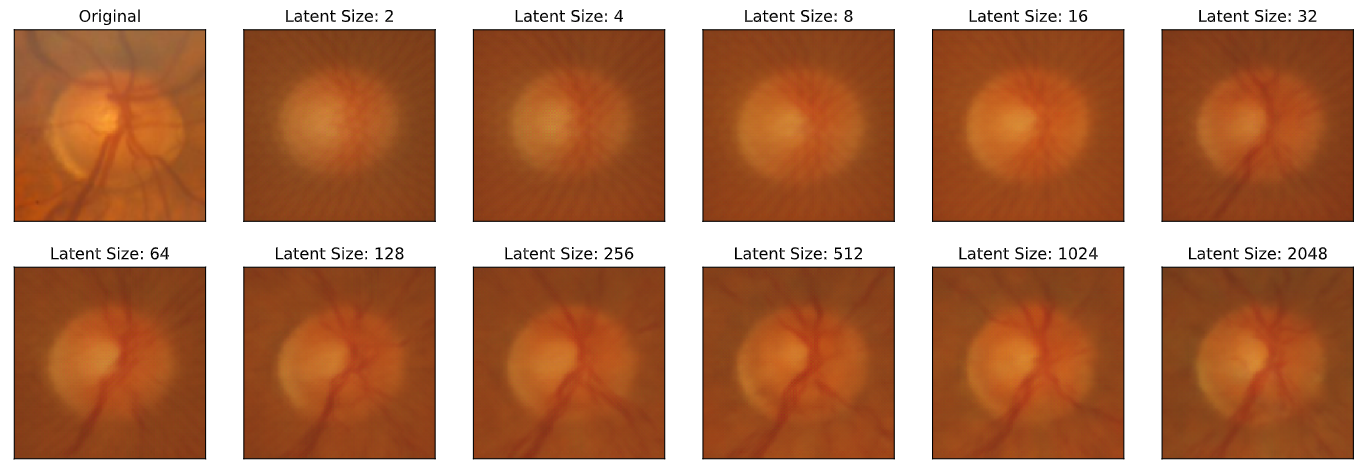}
\caption{Original followed by reconstruction images for normal eye optic disc obtained from trained DFC-VAE at different latent space sizes. The image shows reconstructions of normal eyes are clearly identifiable at latent space sizes $nl \geq 128$. \label{fig:recnorm}}
\end{figure}

\begin{figure}[H]
\centering
\includegraphics[width=0.35\textwidth]{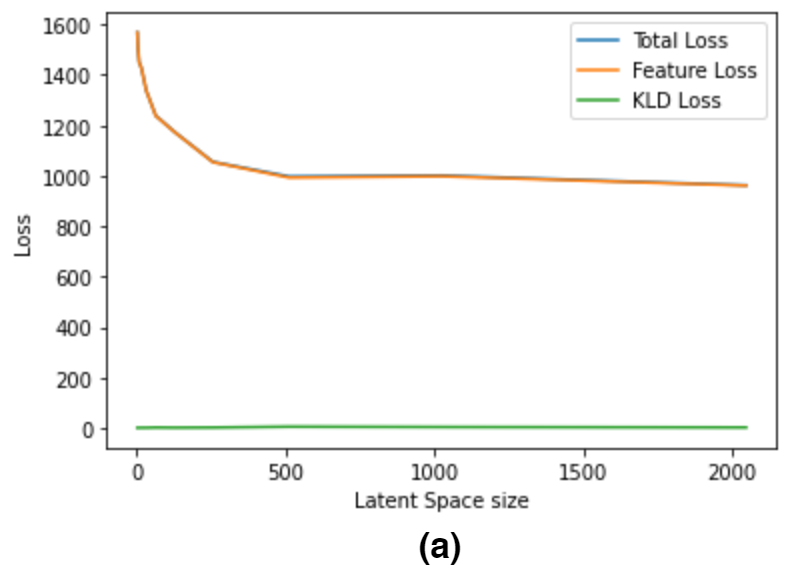}
\hspace{0.2cm}
\includegraphics[width=0.35\textwidth]{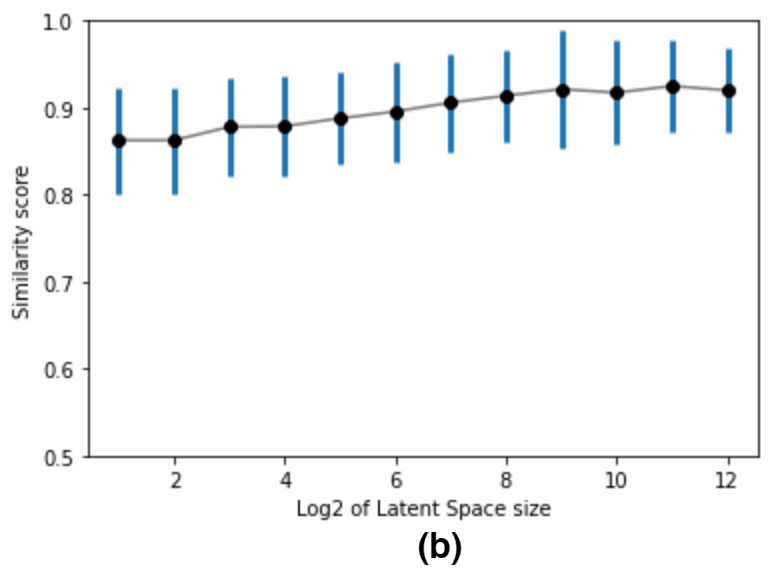}
\caption{Plots for (\textbf{a}) Total, Feature and KL Divergence Loss and (\textbf{b}) Structural similarity score between original and reconstruction images obtained from trained DFC-VAE at different latent space sizes.  \label{fig:eval}}
\end{figure}

The DFC-VAE was trained for eleven different latent space sizes $2^1, 2^2, \dots , 2^{11}$ to identify the bottleneck in learning glaucoma attributes in optic disc images. DFC-VAE with the lowest validation loss was saved for each latent space size and evaluated both qualitatively and quantitatively. Qualitative analysis was done using a visual examination of the reconstructed images. The visual inspection was followed by quantitative analysis by plotting the validation loss and the mean similarity scores between the reconstructed image and the input image at different latent space sizes. The reconstructed images are shown in Figure~\ref{fig:recglau} and Figure~\ref{fig:recnorm} and the plots are shown in Figure~\ref{fig:eval}, which indicates that the DFC-VAE model at appropriate latent space sizes can learn both local and global spatial attributes of the input image. Sample reconstructions of the best DFC-VAE model of each latent space size were reviewed by experts who agreed with latent space size of nl = 128 being the optimal bottleneck at which the DFC-VAE captures most of the relevant glaucoma attributes in optic disc. We also observed that the quality of the reconstructions improved as the latent space size increased.

\subsection{Image Reconstructions}
\label{sec:imgrec}
\begin{figure}[H]
\centering
\includegraphics[width=0.30\textwidth]{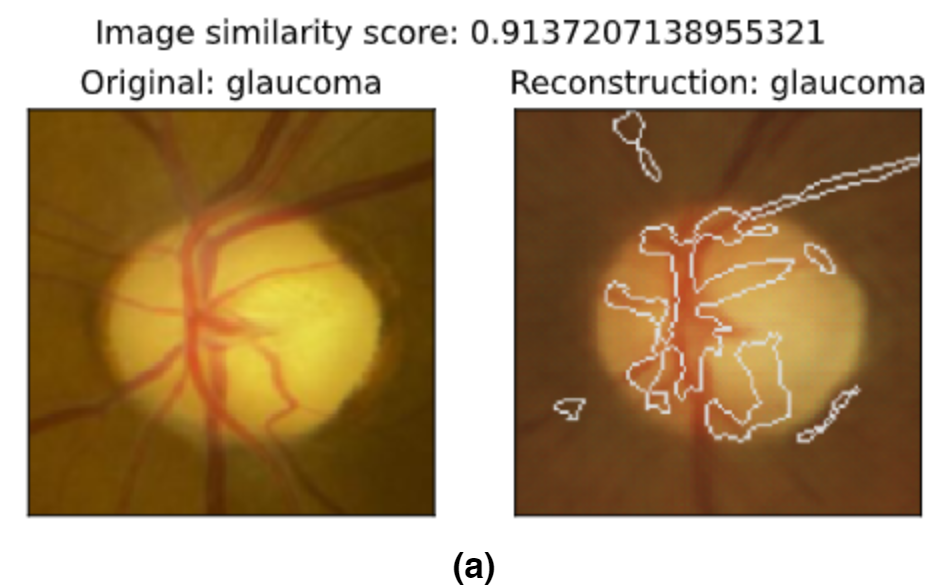}
\hspace{0.2cm}
\includegraphics[width=0.30\textwidth]{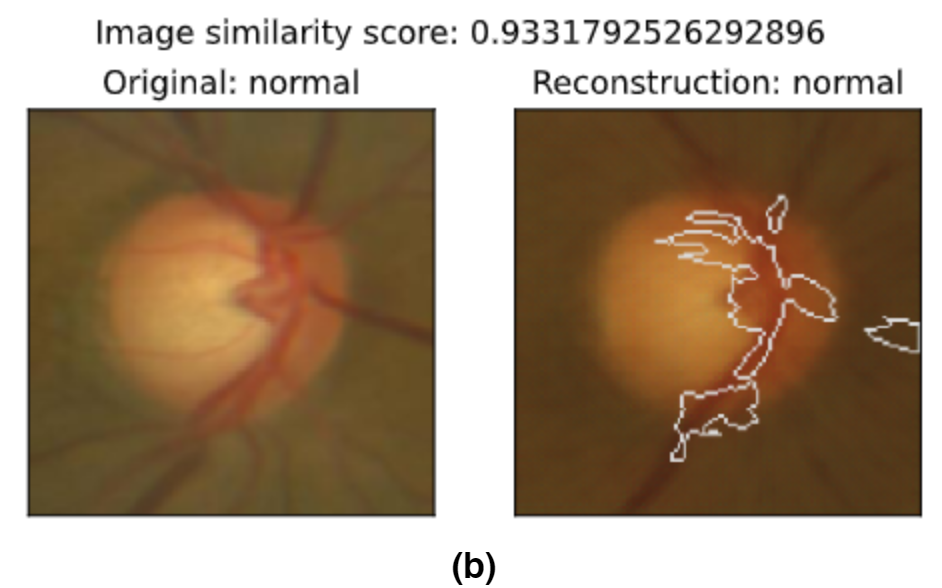}
\includegraphics[width=0.30\textwidth]{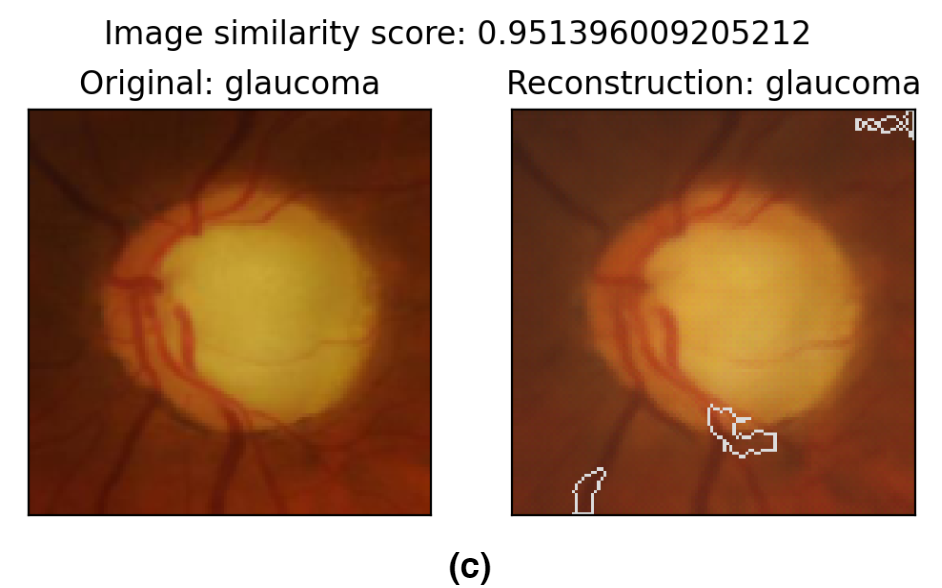}
\hspace{0.2cm}
\includegraphics[width=0.30\textwidth]{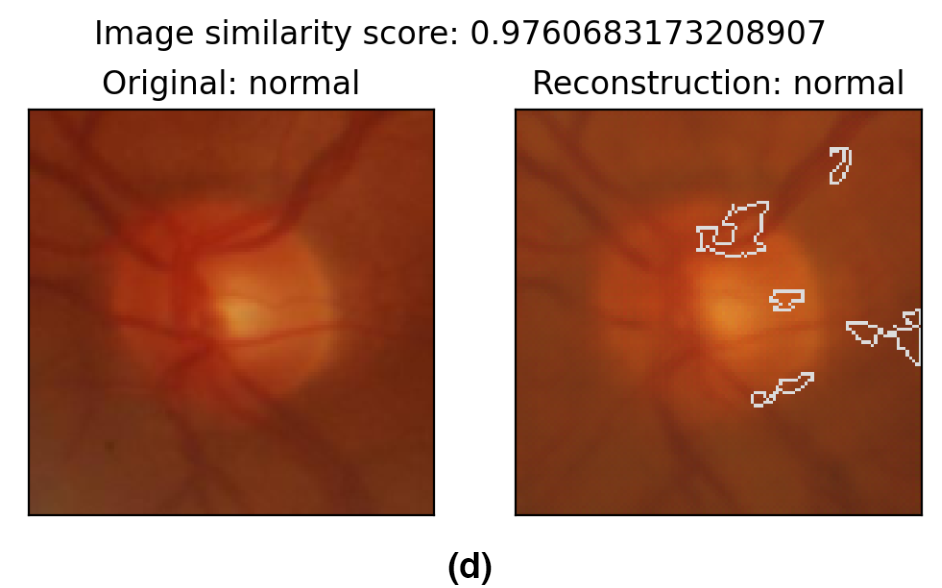}
\caption{Original - Reconstruction image pairs with differences marked in white masks along with similarity scores for (\textbf{a}) Glaucoma eye and (\textbf{b}) Normal eye obtained by DFC-VAE with latent space size $nl=128$, (\textbf{c}) Glaucoma eye and (\textbf{d}) Normal eye obtained by DFC-VAE with latent space size $nl=2048$. \label{fig:diff}}
\end{figure}

For a holistic understanding of features captured by the VAE, the difference between reconstruction and input images were analyzed for the optimal and the largest latent space size (nl = 128 and 2048). This analysis was done by measuring the structural similarity index \cite{ssim} and generating image masks that point out the difference between the two images. The image masks are shown in Figure~\ref{fig:diff}, from which it is evident that the structural similarity increases as the latent space size increases. The image masks also show that all the reconstructions have correctly approximated glaucoma attributes in the reconstructions since observable image difference is on regions which glaucoma is least likely to affect. Most differences occur at the blood vessels both inside and outside the optic disc. This difference was anticipated since VAEs approximate data from a Gaussian distribution. The higher quality of the optic disc images, along with the clear edges of the blood vessels in the optic disc, made it difficult for the VAE to encode it in the latent space.

\subsection{Feature selection and Clustering}
\begin{figure}[H]
\centering
\includegraphics[width=0.22\textwidth]{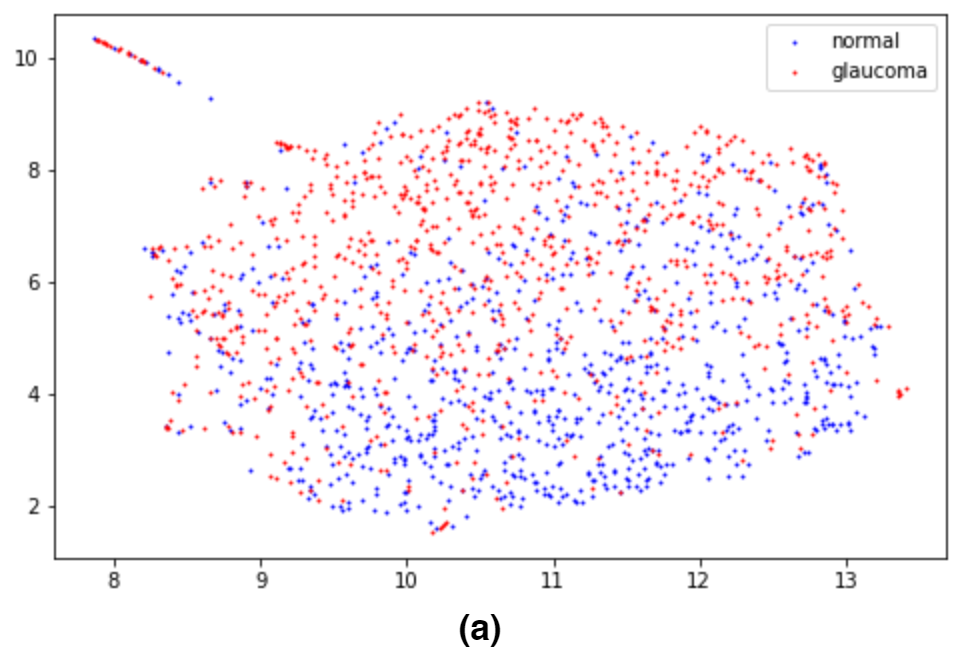}
\hspace{0.2cm}
\includegraphics[width=0.22\textwidth]{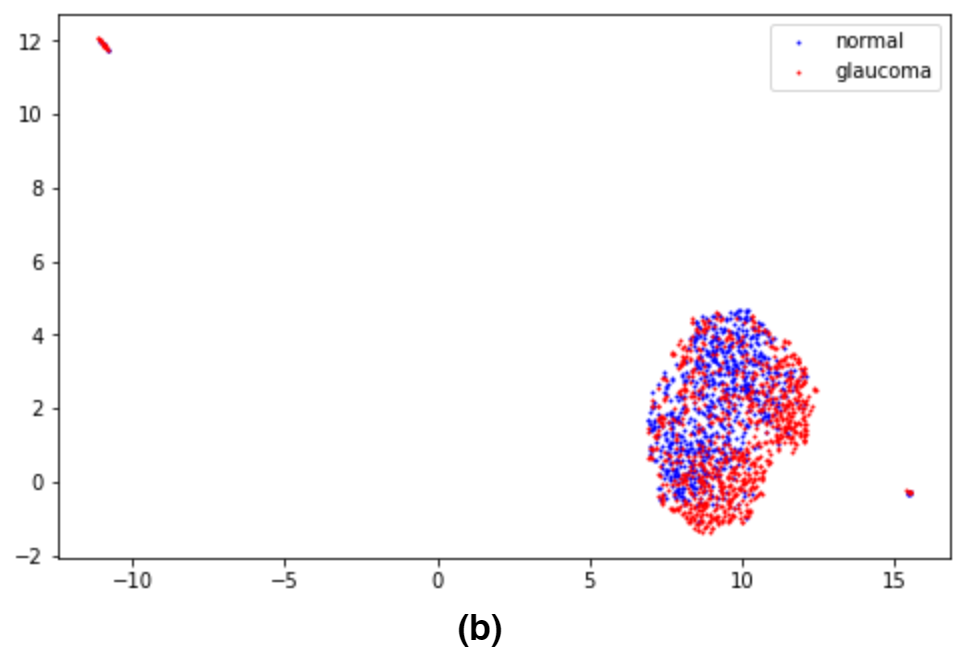}
\hspace{0.2cm}
\includegraphics[width=0.22\textwidth]{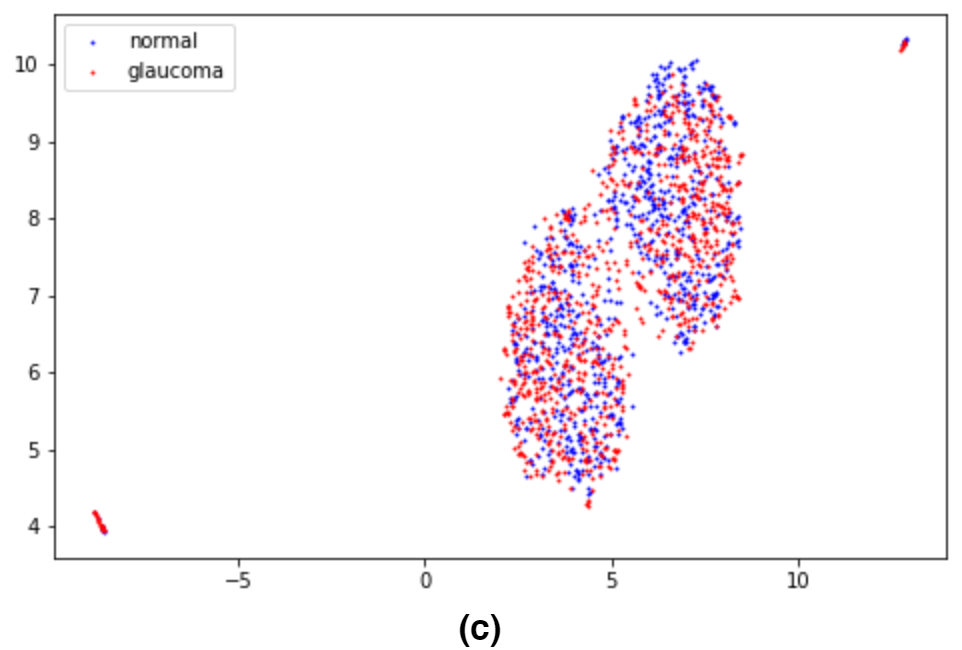}
\includegraphics[width=0.22\textwidth]{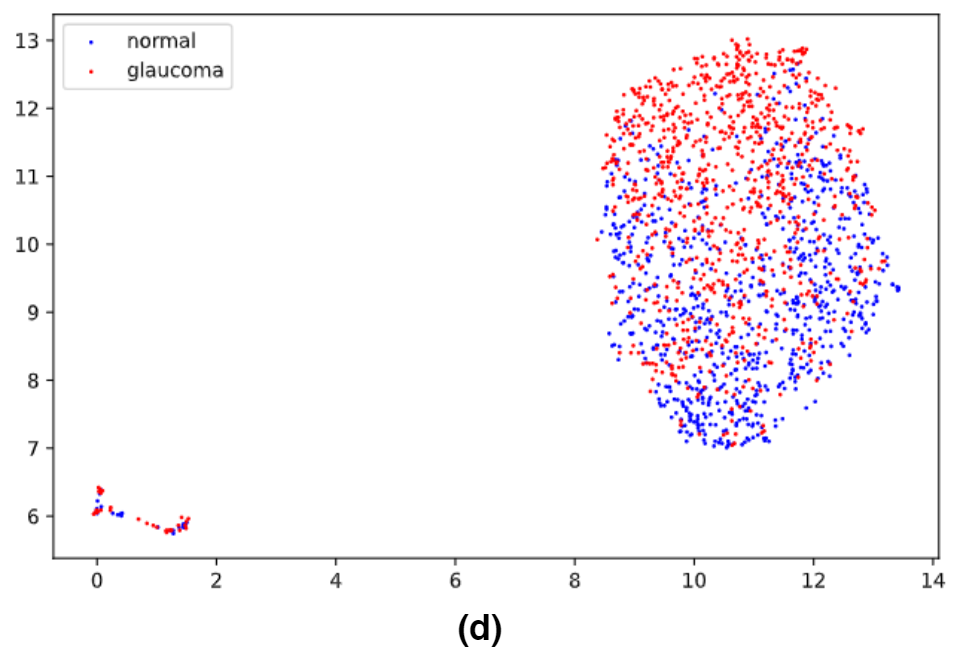}
\hspace{0.2cm}
\includegraphics[width=0.22\textwidth]{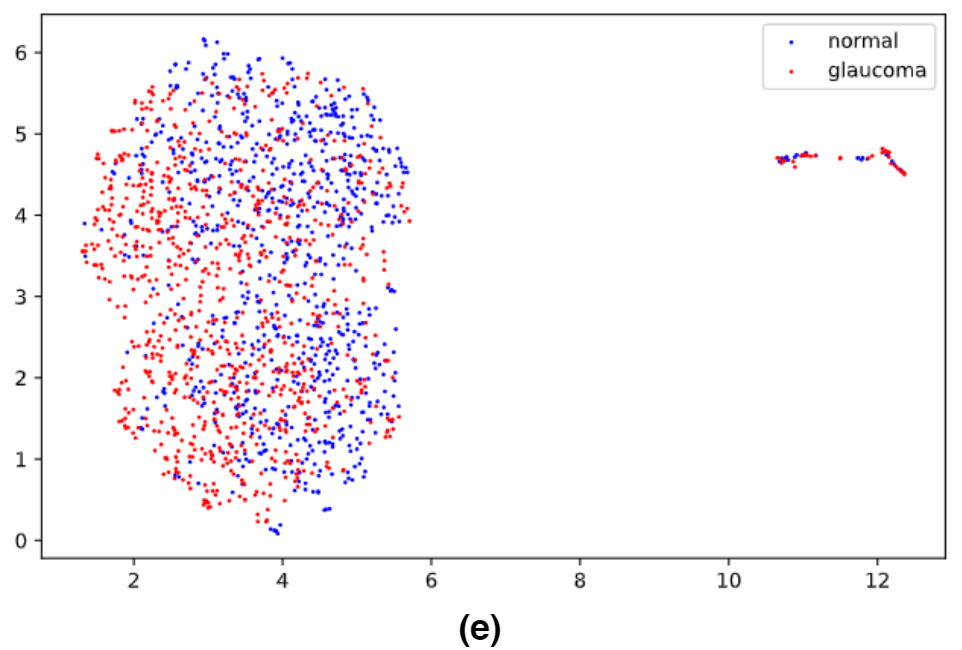}
\hspace{0.2cm}
\includegraphics[width=0.22\textwidth]{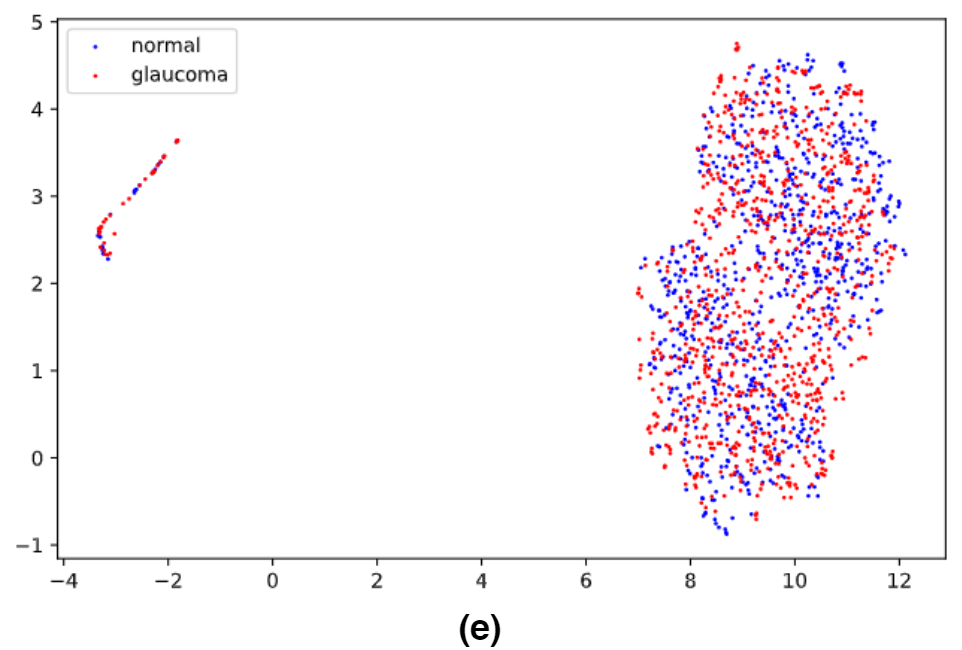}
\caption{2-Dimensional UMAP cluster plots obtained for (\textbf{a}) Top-10 latent space features, (\textbf{b}) Top-50 latent space features and (\textbf{c}) All latent space features obtained by DFC-VAE with latent space size $nl = 128$, and (\textbf{d}) Top-160 latent space features, (\textbf{e}) Top-800 latent space features and (\textbf{f}) All latent space features obtained by DFC-VAE with latent space size $nl = 2048$.  \label{fig:cluster}}
\end{figure}

VAEs encode a complex data distribution to a low-dimensional representation with a probabilistic understanding of a learned attribute. In our study, we were interested if the DFC-VAE learned feature attributes in images related to glaucoma. We did this by observing the separation of the latent representation discovered by DFC-VAE using UMAP clustering. As discussed in methods, we obtained top-10, top-50, and all latent features for latent space size nl=128 and top-160, top-800, and all latent features for latent space size nl=2048 (keeping top-k ratios $(k/nl)$ the same for both latent space sizes) and plot them using UMAP dimensionality reduction and clustering. Figures~\ref{fig:cluster}a-c and Figures~\ref{fig:cluster}d-f shows that the normal-glaucoma cluster separation increases at first then starts decreasing to an inseparable normal-glaucoma cluster. This clustering effect suggests relevant glaucoma attributes make up only a particular portion of the latent space for all latent space sizes. Since cluster separation shifts from normal-glaucoma separation to some other unknown cluster separation (out of the scope of our study), we can conclude that glaucoma attributes present in the input images might not be as dominating as other irrelevant features. This shift bolsters the fact that glaucoma detection is difficult in usual settings (human grading) but can be improved with deep learning methods. In any case, we can see that a clear separation exists for normal-glaucoma eye clustering, which means the DFC-VAE has learned significant glaucoma attributes from the high dimensional, complex optic disc data.

\subsection{Glaucoma Classification}
\begin{figure}[H]
\centering
\includegraphics[width=0.45\textwidth]{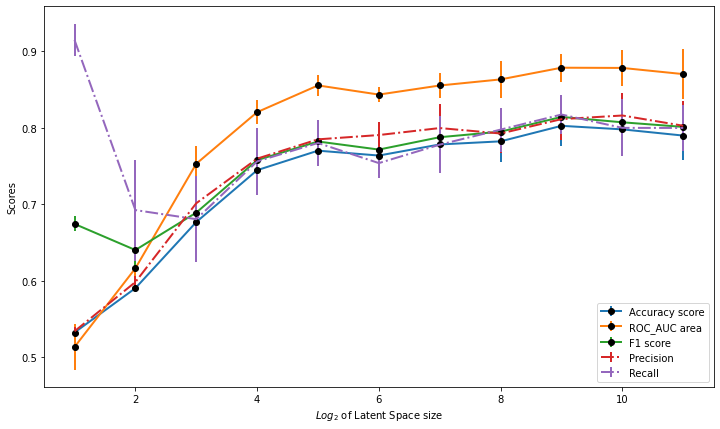}
\caption{5-Fold Cross Validation Performance Trends for optimal SVC classifier model trained with DFC-VAE latent representations obtained at different latent space sizes. \label{fig:perf}}
\end{figure}

We evaluated glaucoma classification based on five performance metrics which provide a quantitative estimate of the DFC-VAEs learnability of relevant features in the optic disc. Initially, a 5-fold cross-validation was done with the SVC classifier to obtain optimal hyperparameters. Training the 5-fold cross-validation reported optimal hyperparameters: C=1, Kernel = Radial Basis Function, and Class Weight = Balanced. Plot (Figure~\ref{fig:perf}) shows and compares the best of all five performance score trends over different latent space sizes for this model. It is apparent that the DFC-VAE latent representations at size $nl\geq 128$ has high performance scores indicating that the DFC-VAE can encode glaucoma attributes from optic disc images.

\begin{figure}[H]
\centering
\includegraphics[width=0.45\textwidth]{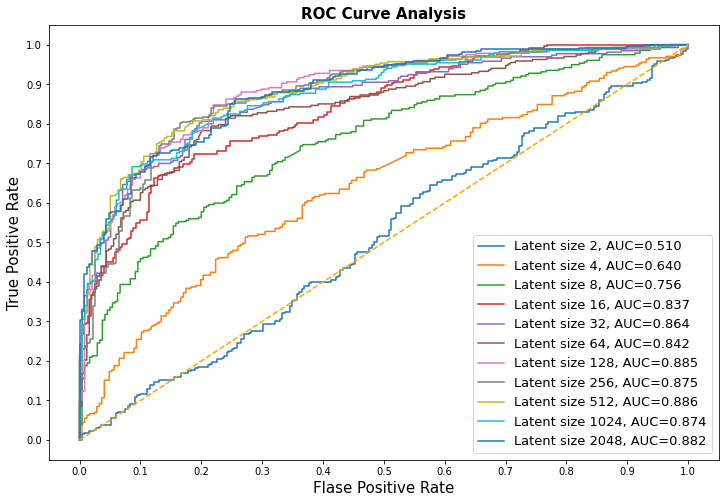}
\caption{ROC curve and AUC scores obtained from Train-Test split based SVC classifier model trained with DFC-VAE latent representations as inputs. \label{fig:roc}}
\end{figure}

The 5-fold cross-validation model was followed by a simple train-test split model with a split ratio of 0.3 and the optimal SVC classifier as the learning algorithm. ROC curves generated for the classification model at each latent space size are shown in Figure~\ref{fig:roc}. Overall the classifier reported an excellent correlation between latent representations and class labels with AUC score 0.885 at latent space size as low as nl = 128.Even latent space size as low as nl = 16 in our method reported good discriminations with an AUC score 0.837.

\section{Discussion}

This paper showed that deep unsupervised learning could effectively learn the underlying glaucoma data distribution of optic disc images without any input labels. The DFC-VAE developed can compress (encode) all high-level features in the optic disc (49152 pixels) effectively in a low dimensional latent space of size as small as nl=128 and can generate (decode) clinically appropriate reconstructions from the latent space. The model validation is done by both qualitative and quantitative analysis of the reconstructions and the latent space. The study included a visual examination of the reconstructions and calculation of structural similarity score. We observed that the latent space of the DFC-VAE could retain most of the glaucoma attributes in the optic disc. This observation is bolstered by (1) feature selection and clustering relevant low-dimensional latent features from the latent space and (2) analyzing the latent representation’s ability to classify eyes with glaucoma for different latent space sizes. A Area under the ROC Curve score of 0.885 was obtained with an SVC classifier with optimal hyperparameters which was trained with DFC-VAE latent representations obtained at size $nl=128$.

A direct comparison of the current study with previous research \cite{diaz2018retinal} shows that the DFC-VAE based glaucoma assessment is superior to the VAE results obtained in the previous study. The reconstructions obtained by the DFC-VAE are qualitatively as sharp and structurally similar to the reconstructions obtained by a similarly trained Deep Convolutional Generative Adversarial Network (DCGAN) \cite{diaz2018retinal}.  This observation assures that the complex data can be projected to a low-dimensional space with efficient probabilistic data distribution while retaining most visual attributes. As shown in section~\ref{sec:imgrec}, most differences in image reconstructions can be observed at regions that have sharp edges, such as small blood vessels. Thus, visual attributes in the optic disc images important in the clinical detection of glaucoma appear to remain intact. In eyes with glaucoma, prominent features such as enlarged cup, retinal nerve fiber layer defects, or rim thinning are identifiable in the DFC-VAE image reconstructions.


Clustering and classification results of the DFC-VAE based latent representations of optic disc images showed agreement with prior research on glaucoma assessment \cite{diaz2018retinal,kanagala2020detection}. Readers should note that although CNN-based supervised deep learning methods exist for glaucoma analysis \cite{detection2021medeiros,detection2021shabbir,Kedarnath2020GlaucomaDU,LEE202186,computer2018hagiwara}, they do not account for the explainability of the attributes of the optic disc, which are responsible for glaucoma. Given that most supervised learning methods are dependent on human graded labels, the learning process becomes blindsided, making glaucoma assessment error-prone. The DFC-VAE approach used in this study learns both relevant and trivial explainable patterns in the data responsible for glaucoma detection. This learnability provides a level of trust and reliability in glaucoma assessment, albeit the classification accuracy might be slightly lower.

Previously, Berchuck et al. showed that latent representations learned by generalized VAEs are superior to global parameters such as MD in estimating the progression and predicting future visual fields in glaucoma \cite{estimating2019berchuck}.  Similar observations were found when an artificial neural network (ANN) based VAE was trained to learn low-dimensional latent representations of SD OCT, where glaucoma classification AUC was found to be higher for low-dimensional latent features than standard RNFL thickness measurements \cite{mariottoni2020clustering}. These results are analogous to the observations obtained in the current study. Since we showed that the DFC-VAE could learn important discernible glaucoma attributes for clinical diagnosis, we can draw parallels between the three studies. Specifically, through the current analysis, we point out the possibility of an integrated, comprehensive analysis of glaucoma detection, development, and progression using multiple data types using DL approaches.

Further, it is to be noted that although the emphasis of the current research is on finding clinically appropriate latent representations of glaucoma, our analysis has also paved the way for the interpretability of deep learning methods in glaucoma detection. The reconstructions obtained by the DFC-VAE can be used to analyze how deep learning methods may fall short during classification tasks. Since the reconstructions are derived from a Gaussian probability distribution, the image differences obtained from the structural similarity between the reconstruction and input may provide hidden insights on glaucoma features.

\section{Conclusions}

In this paper, we developed a DFC-VAE based unsupervised learning approach to extract information from optic disc images. The DFC-VAE model learns to identify and compress glaucoma attributes into latent representations from the complex data distribution. Validation of the DFC-VAE is done by comparing the structural similarity of reconstructions with the original input. This process is repeated for different latent space sizes to obtain a bottleneck latent space that can accurately capture all relevant glaucoma attributes. The feature attributes extracted from the model are then evaluated by UMAP cluster analysis and SVC classification, showing an excellent correlation between latent representations and class labels. Due to the probabilistic nature of the latent representations, our future work will be to extend the current research to analyze glaucoma progression in optic disc images and compare it with other research in similar domains.


\vspace{6pt} 



\authorcontributions{Conceptualization, S.M., A.A.J. and F.A.M.; methodology, S.M.; software, S.M.; validation, S.M., A.A.J. and F.A.M.; formal analysis, S.M.; investigation, S.M.; resources, S.M. and F.A.M.; data curation, A.A.J. and F.A.M; writing---original draft preparation, S.M.; writing---review and editing, A.A.J. and F.A.M.; visualization, S.M.; supervision, F.A.M.; project administration, F.A.M.; funding acquisition, F.A.M. All authors have read and agreed to the published version of the manuscript.}

\funding{Supported in part by National Institutes of Health/National Eye Institute grant EY029885 and EY031898 (FAM). The funding organization had no role in the design or conduct of this research.}

\institutionalreview{The Duke University Institutional Review Board approved this study with a waiver of informed consent due to the retrospective nature of this work.}

\informedconsent{Informed consent was obtained from all subjects involved in the study.}


\dataavailability{The datasets generated during and/or analyzed during the current study are available from the corresponding author on reasonable request.} 

\acknowledgments{None.}

\conflictsofinterest{S.M.: none. A.A.J.: none. F.A.M.: Aeri Pharmaceuticals (C);  Allergan (C, F), Annexon (C); Biogen (C); Carl Zeiss Meditec (C, F), Galimedix (C); Google Inc. (F); Heidelberg Engineering (F), IDx (C); nGoggle Inc. (P), Novartis (F); Stealth Biotherapeutics (C); Reichert (C, F).} 

\reftitle{References}


\externalbibliography{yes}
\bibliography{ext_bib.bib}


%


\end{document}